\documentclass[12pt,oneside]{article}
\pdfoutput=1
\usepackage[pdftex]{graphicx}
\usepackage{amssymb}
\usepackage{amsmath}
\usepackage{supertabular}
\usepackage{fancyhdr}
\begin{document}

\newcommand\Block[2]{%
\begin{minipage}[c][.27\textheight][t]{0.5\textwidth}
#1\par #2
\end{minipage}%
}

\begin{titlepage}
\vspace*{1.0cm}

\begin{center}
{\Large \bf
Investigation of the Galactic Magnetic Field \\
\vspace*{5mm}
with Ultra-High Energy Cosmic Rays}
\vskip2cm

Martin Erdmann, Gero M\"uller, Martin Urban
\end{center}

\vskip1cm

\begin{center}
\begin{minipage}[]{.875\textwidth}
\begin{small}
\noindent
Phys. Institute 3A, RWTH Aachen University, D-52056 Aachen, Germany\\
E-mail: erdmann@physik.rwth-aachen.de
\end{small}

\vskip2cm

\begin{small}
{\bf Abstract:}
We present a method to correct for deflections of 
ultra-high energy cosmic rays in the galactic magnetic field.
We perform these corrections by simulating the expected arrival directions of protons 
using a parameterization of the field derived from Faraday rotation and synchrotron 
emission measurements.
To evaluate the method we introduce a simulated astrophysical scenario and
two observables designed for testing cosmic ray deflections.
We show that protons can be identified by taking advantage of the galactic magnetic field pattern.
Consequently, cosmic ray deflection in the galactic field can be verified experimentally.
The method also enables searches for directional correlations of cosmic rays with source 
candidates.
\end{small}
\end{minipage}
\end{center}

\end{titlepage}

\newpage

\section{Introduction}

The magnetic field of our galaxy is presumed to cause substantial deflections of ultra-high 
energy cosmic rays prior to their observation. 
To understand the deflections, the directional dependence of the field pattern and the 
directional variations in the magnitude of the field are most relevant.
\par
Explicit predictions for deflections depending on the cosmic ray arrival direction, 
its energy and charge can be obtained through recent parameterizations of the shape 
and magnitude of the galactic magnetic field
\cite{Pshirkov2011, Jansson2012a, Jansson2012b, Pshirkov2013, Beck2014}.
They are based on $40,000$ Faraday rotation and synchrotron emission measurements.
Exploring cosmic ray deflections thus constitutes an independent test of the field 
parameterizations, and can in addition be related to obtaining information on the 
origin of cosmic rays.
\par
In this contribution we present an analysis method to investigate cosmic ray 
deflections in magnetic fields.
The method is based on propagating protons through the field to obtain their expected 
arrival directions.
Owing to the forward simulation technique we can test the validity of a field 
parameterization without imposing potentially unphysical conditions on measured cosmic rays.
\par
Our analysis concept consists of testing a combined model with three assumptions: 
The first assumption is that coincident directions of cosmic rays and source candidates 
exist outside of our galaxy, before the cosmic rays are deflected in the galactic magnetic field. 
The second aspect of the model is that the above parameterizations of the galactic 
magnetic field reflect the field characteristics and magnitude as realized in our galaxy. 
The third assumption is that the magnetic field is sufficiently strong to 
separate protons from nuclei through small angular deflections.
\par
To assess the method we simulate an astrophysical scenario that we call `data'
throughout this contribution.
To generate these data we use the cosmic ray propagation program CRPropa version $3$ 
\cite{Batista2014}.
As sources of cosmic rays have not been identified yet, we make the following choice
for the simulated data:
We assume active galactic nuclei (AGN) within the GZK horizon \cite{GZK} to be source candidates,
and use neutrinos as navigators for selecting source directions.
\par
This contribution is structured as follows. 
We first explain the simulated astrophysical scenario and the corresponding simulated data set.
We then present the method of calculating the expected arrival directions of cosmic
rays after traversing the magnetic field. 
We analyze the angular distances between the observed cosmic ray arrival directions 
and the expected arrival directions. 
We also quantify the probability of finding several cosmic rays associated 
to single AGNs. 
We then study the effect of alternative model assumptions when analyzing the simulated data.
Finally we present our conclusions.

\section{Astrophysical scenario}

Here we assume that a subset of active galactic nuclei (AGN) can be considered as sources
of cosmic rays.
As navigators to select these AGN sources experimentally we use the arrival directions of
$24$ high energy neutrinos published by the IceCube Collaboration \cite{icecube2014}.
These neutrinos have energies above $E=40$ TeV, and were not assigned strong evidence 
for resulting from atmospheric background. 
\par
\begin{figure}[hbt]
\begin{minipage}[]{.6\textwidth}
\flushleft{a)}
\end{minipage}
\begin{minipage}[]{.4\textwidth}
\flushleft{b)}
\end{minipage}
\begin{minipage}[]{.6\textwidth}
{\centering
     \includegraphics[width=\textwidth]{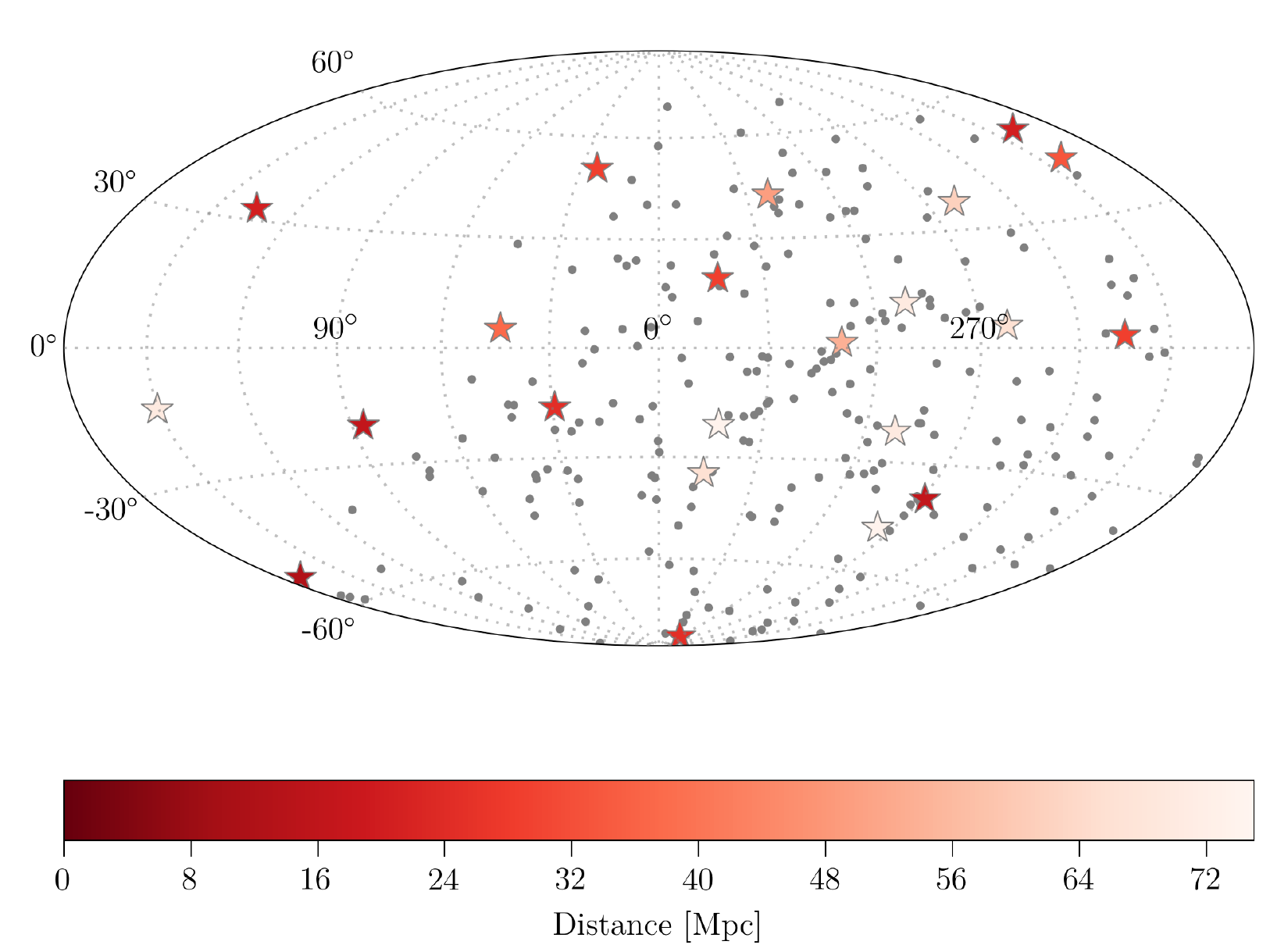}
}
\end{minipage}
\begin{minipage}[]{.4\textwidth}
\vspace*{-2cm}
{
\hspace*{1.5cm}
     \includegraphics[width=0.7\textwidth]{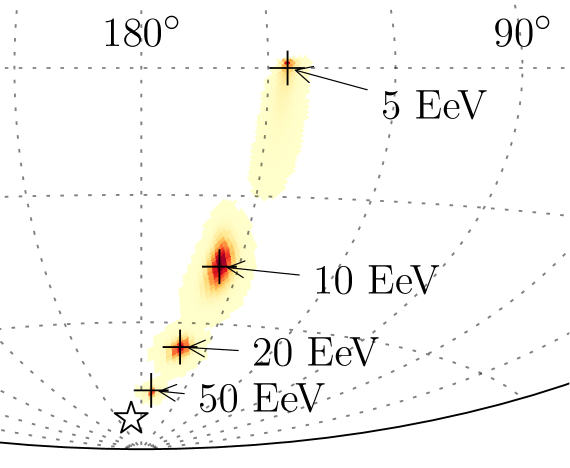}
}
\end{minipage}
\caption{a) Simulated astrophysical scenario. 
The star symbols denote AGNs with the color corresponding to their distance.
The circular symbols represent the cosmic rays ($7\%$ proton signal, 
$93\%$ background contributions).
b) Simulations of the expected arrival direction of protons after traversing the 
galactic magnetic field.
The star symbol denotes the initial direction outside the galaxy, the cross symbols 
show the expected arrival directions of protons with different energies, 
and the color code gives relative probability distributions.
}
\label{fig:scenario}
\end{figure}

For the AGNs we use the VCV catalogue \cite{vcv12} and consider AGNs with distances $z<0.018$.
This value is motivated by the analysis performed in \cite{AUGER} and provides sources within 
the GZK horizon.
With the criterion of selecting the nearest AGNs of the $24$ neutrinos we obtain $22$ AGNs
where two neutrinos have the same AGN.
The AGN sources are shown by the star symbols in Fig. \ref{fig:scenario}a, with the color 
code indicating the AGN distances.
\par
Our aim is to obtain a simulated cosmic ray data set corresponding to $N=231$ ultra-high 
energy cosmic rays with distributions similar to the one published by the Pierre Auger 
Collaboration \cite{Aab2015a}.
For this we combine $10\%$ of an astrophysical scenario simulated with the CRPropa program
\cite{Batista2014} with $90\%$ isotropic cosmic rays following the geometrical acceptance of the 
observatory \cite{Sommers2000}.
\par
For the $10\%$ contribution of the astrophysical scenario we generate $10^{7}$ cosmic rays 
at each of the $22$ selected AGN sources.
As the initial composition we use a flat distribution of nuclei with 
charges between $Z=1,...,26$.
Their minimum energies are $E=50$ EeV, and their maximum energies correspond to their
rigidities $Z\times E_{max}$ with $E_{max}=500$ EeV.
\par
Upon arrival at a $0.5$ Mpc observer sphere around our galaxy
we use galactic magnetic field lenses to project the arriving cosmic rays onto Earth.
The lenses consist of matrices based on the HEALPix format \cite{Gorski2005} 
where we divide the sphere into $49,152$ equally sized pixels of approximately $1\;\deg$.
The matrices contain the probability of a cosmic ray entering the galaxy
with energy $E$ at direction $i$ to be observed in direction $j$.
The technical details of the lenses and their production are outlined in \cite{Bretz2014}.
The lenses used in this contribution have been calculated with the CRPropa program and the 
JF12 parameterization of the galactic magnetic field \cite{Jansson2012a}.
\par
We then select cosmic rays according to the geometrical acceptance of the observatory 
\cite{Sommers2000}, and an energy distribution corresponding to the measured energy 
spectrum \cite{AUGER_energy}.
Of these cosmic rays we randomly select the 
$10\%$ contribution of cosmic rays arriving from the AGN sources
where we require $70\%$ to be protons.
The total sample of the $231$ simulated cosmic rays therefore 
contains $7\%$ proton signal and $93\%$ background
(grey circular symbols in Fig. \ref{fig:scenario}a).

\section{Analysis method}

To investigate the galactic magnetic field using the cosmic ray data set described in 
the previous section, we simulate their expected arrival directions assuming that all cosmic rays 
are protons.
We start these proton simulations with the same set of the $22$ AGNs mentioned above.
\par
To take into account effects of extragalactic magnetic fields we apply a Fisher probability 
distribution centered at each AGN direction with a concentration parameter $\kappa$ depending 
on the AGN distance and the cosmic ray energy.
For an AGN at distance $D=10$ Mpc and a proton with $E=52$ EeV the angular spread 
amounts to $\sigma=1/\sqrt{\kappa}=6$ deg.
\par
This probability distribution is then projected onto the Earth using the corresponding 
magnetic lens described above.
To obtain a single direction for which the arrival probability is greatest, 
we calculate the radius $r_{50}$ containing $50\%$ of the arrival probabilities. 
We select the pixel with the smallest radius $r_{50}$  and use the center of this pixel 
as the expected arrival direction of the proton.
\par
The procedure for calculating the expected arrival direction of protons with energies 
$E=5, 10, 20,$ $50$ EeV is visualized exemplarily in Fig. \ref{fig:scenario}b.
The star symbol denotes the initial direction outside the galaxy. 
Note that the direction of the lowest energy proton also corresponds to the deflection 
of an ionized Neon nucleus ($Z=10$) with energy $E=50$ EeV. 
This implies that - in directions with a sufficiently strong magnetic field - protons in 
the cosmic ray data can be identified to some extent by small angular distances to the 
expected arrival directions.

\section{Angular distance and clustering strength}

\begin{figure}[hbt]
\Block{a)}{
    %\vspace{-5ex}
    \centering
    \includegraphics[width=0.70\textwidth]{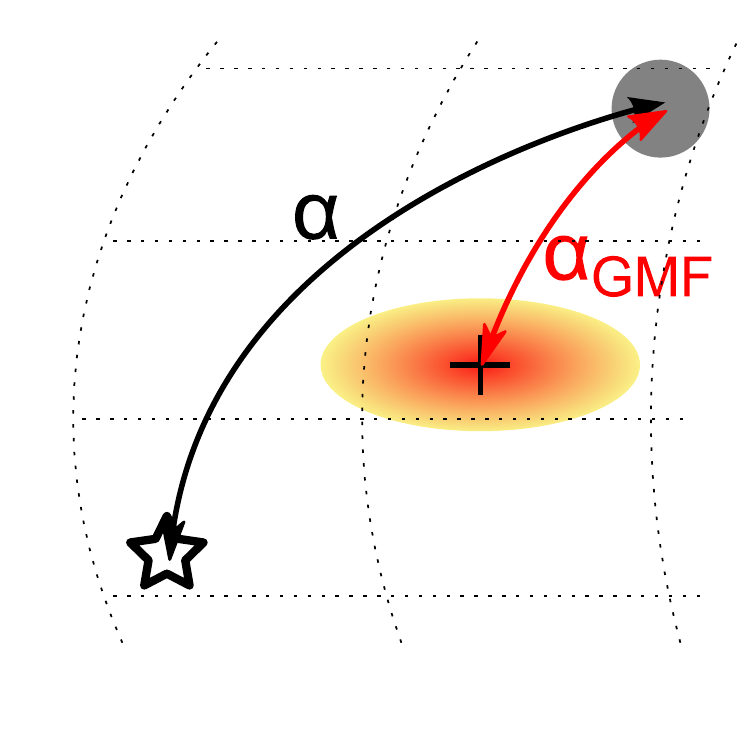}
}
\Block{b)}{
    %\vspace{-5ex}
    \centering
    \includegraphics[width=0.80\textwidth]{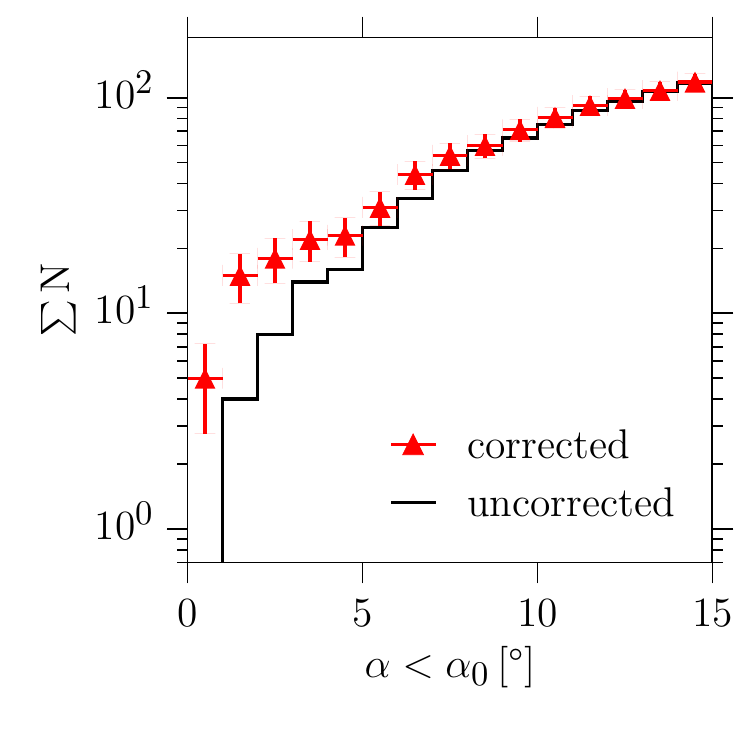}
}
% \par\nointerlineskip\noindent
    \\[5ex]
\Block{c)}{
    %\vspace{-5ex}
    \centering
    \includegraphics[width=0.80\textwidth]{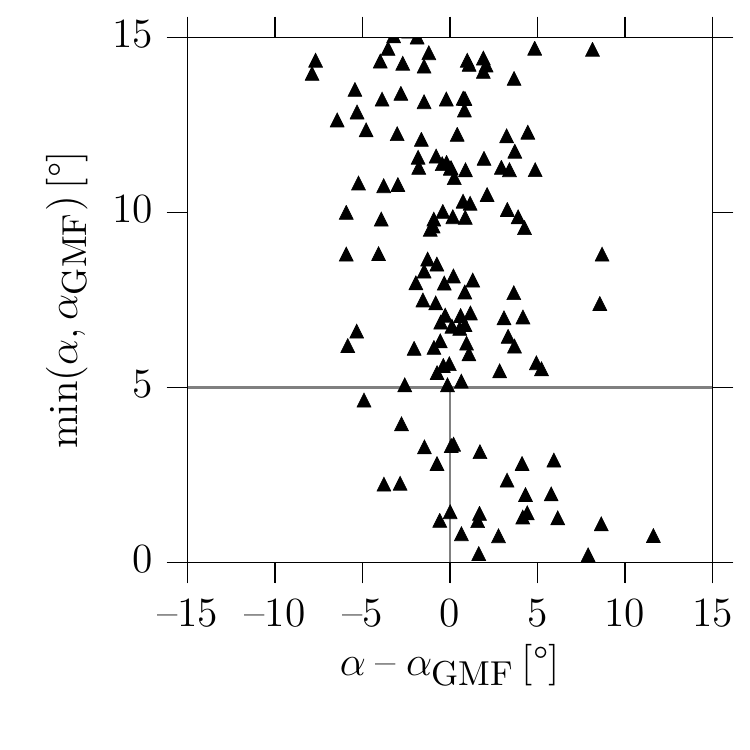}
}
\Block{d)}{
    %\vspace{-5ex}
    \centering
    \includegraphics[width=0.80\textwidth]{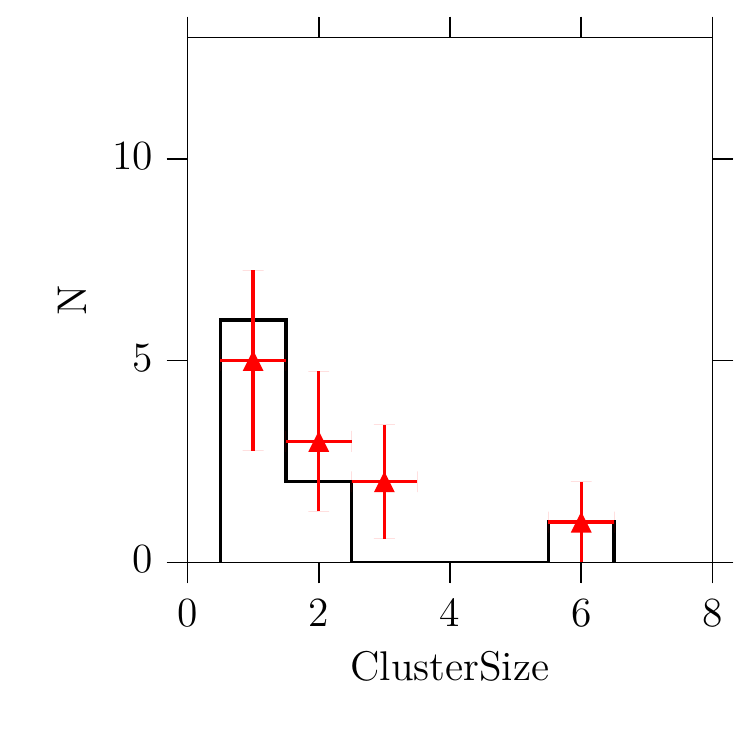}
}
    \\[5ex]
\caption{
a) Angular distances of a cosmic ray to the nearest AGN direction and to the
expected arrival direction assuming protons.
b) Cumulated number of cosmic rays associated with AGN arrival directions
with (symbols) and without magnetic field corrections (histogram) as a function of
the maximum angular distance $\alpha_\circ$.
c) Change in the angular distance before and after applying the magnetic field corrections.
The vertical line separates reduced ($>0$) and enlarged ($<0$) angular distances.
The vertical axis shows the smallest angular distance, whereas the horizontal line emphasizes 
the region below $5$ deg.
d) Frequencies of the cluster sizes $m$ for the maximum angular distance $5$ deg.
}
\label{fig:observables}
\end{figure}

In Fig.\ref{fig:observables}a we visualize two angular distances used in this analysis. 
The angle $\alpha$ denotes the angular distance between the measured arrival directions 
of a cosmic ray and its nearest AGN.
The angle $\alpha_{\rm GMF}$ denotes the angular distance between the measured cosmic ray
and the expected arrival direction of a proton with an AGN-coincident direction 
outside the galaxy.
\par
In Fig.\ref{fig:observables}b we show the cumulated number of cosmic rays arriving within a distance 
of angle $\alpha_\circ$ to their nearest AGNs. 
The triangular symbols present the angular distances between the measured cosmic rays 
and the expected arrival directions for protons including the magnetic field corrections. 
The histogram denotes the uncorrected angular distances between cosmic rays and AGNs. 
Below a few degrees, more cosmic rays appear near AGN directions when including the field corrections.
These small angular distances are consistent with our simulations of the protons, which supports
the above-mentioned third assumption of our model that protons in the cosmic ray sample can be
identified to some extent by exploiting the magnetic field deflections.
\par
In Fig.\ref{fig:observables}c we present the change in the angular distances $\alpha-\alpha_{\rm GMF}$ 
without and with magnetic field corrections. 
At positive values the cosmic rays come closer to the expected arrival directions, 
while at negative values the cosmic rays are nearer to the AGNs without field 
corrections. 
In the region with small angles we find more cosmic rays with improved angular distances when applying 
the field corrections compared to events for which the uncorrected distance is smaller.
\par
We quantify the change in the angular distances by the asymmetry
\begin{equation}
    \label{eq:asymmetry}
    A \equiv 2 \; \frac{N(\alpha>\alpha_{\rm GMF}) - N(\alpha<\alpha_{\rm GMF})}
                       {N(\alpha>\alpha_{\rm GMF}) + N(\alpha<\alpha_{\rm GMF})}
\end{equation}
which can take on any value from $-2$ to $2$.
Using e.g. a maximum angle of $\alpha_\circ=5$ deg the angular asymmetry in the data is found to be 
$A=0.96$.
This positive value is a measure of the overall improvement in the angular distances between 
cosmic rays and AGNs when applying the field corrections.

We also investigate clustering of cosmic rays with AGN directions. 
In Fig.\ref{fig:observables}d we present the frequencies of the cluster sizes $m$ where we count associations 
of cosmic rays with AGNs within $5$ deg angular distance.
This yields configurations containing singlet, doublet, triplet, and sextet clusters ($m=1, 2, 3, 6$). 
The triangular symbols represent the data when including magnetic field corrections, 
and the histogram without them.

To quantify the observed clustering strength we use the multinomial probability distribution:
\begin{equation}
P(n_1,..., n_{22}; N-N_{hit}) = 
\frac{N!}{n_1!... n_{22}! \;(N-N_{hit})!}\;\; p_1^{n_1}\; ...\; p_{22}^{n_{22}}\;\; (1-p_{iso})^{N-N_{hit}}
\label{eq:multinomial}
\end{equation}
The value $P$ describes the expected level of trivial clustering between the $N=231$ cosmic rays, 
and the $22$ AGNs where the latter are distinguished by identifiers. 
AGN $i$ has an average hit probability of $p_i$, and was correlated with $n_i$ cosmic rays. 
The total number of cosmic rays associated with one of the AGNs is $N_{hit}=\sum_i n_i$.
The remaining $N-N_{hit}$ cosmic rays without AGN correlations had a no-hit 
probability of $(1-p_{iso})$. Summing the hit probabilities for the AGNs at 
their nominal arrival directions for angular distances below $5$ deg gives $p_{iso}=5.3\%$. 
This includes the geometric acceptance of the Pierre Auger Observatory \cite{Sommers2000}. 
When including the magnetic field corrections a slight energy dependence is observed where
the probability increases on average to $p_{iso}=5.7\%$.
\par
When applying the magnetic field corrections to the data, the logarithm of the 
multinomial probability (\ref{eq:multinomial}) amounts to $\log_{10}(P_{\rm GMF})=-15.3$.
Without the field corrections the level of clustering is smaller and results in 
$\log_{10}(P)=-11.5$.
The change in the clustering strength thus amounts to $\log_{10}(P_{\rm GMF})-\log_{10}(P)=-3.8$.

\section{Analyses using alternative assumptions}

\begin{figure}[hbt]
\Block{a)}{
    %\vspace{-5ex}
    \centering
    \includegraphics[width=0.80\textwidth]{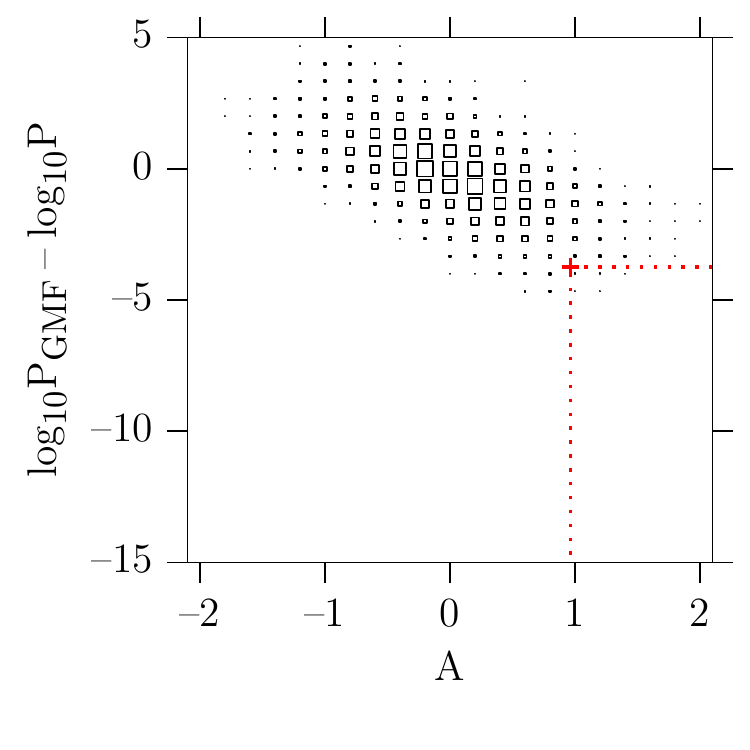}
}
\Block{b)}{
    %\vspace{-5ex}
    \centering
    \includegraphics[width=0.80\textwidth]{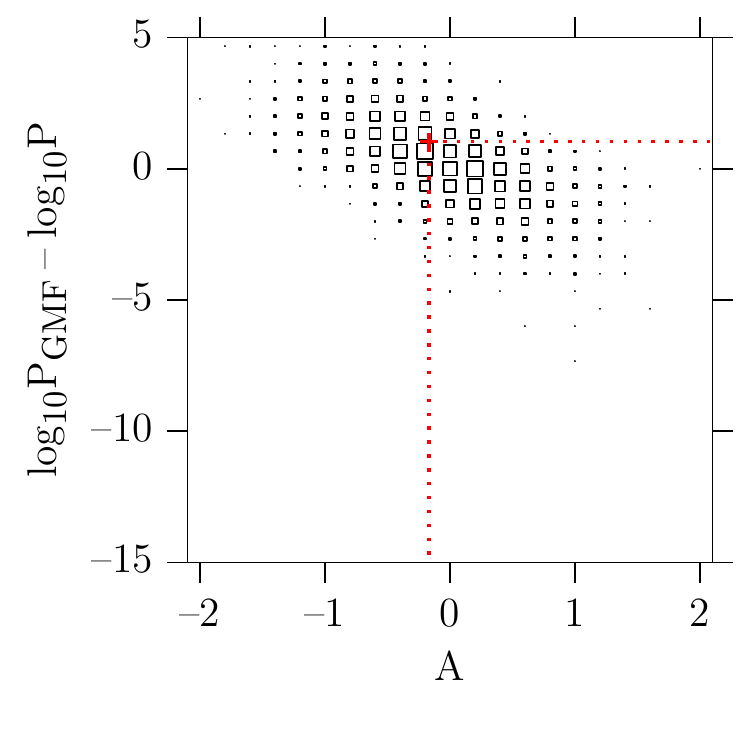}
}
    \\[5ex]
\Block{c)}{
    %\vspace{-5ex}
    \centering
    \includegraphics[width=0.80\textwidth]{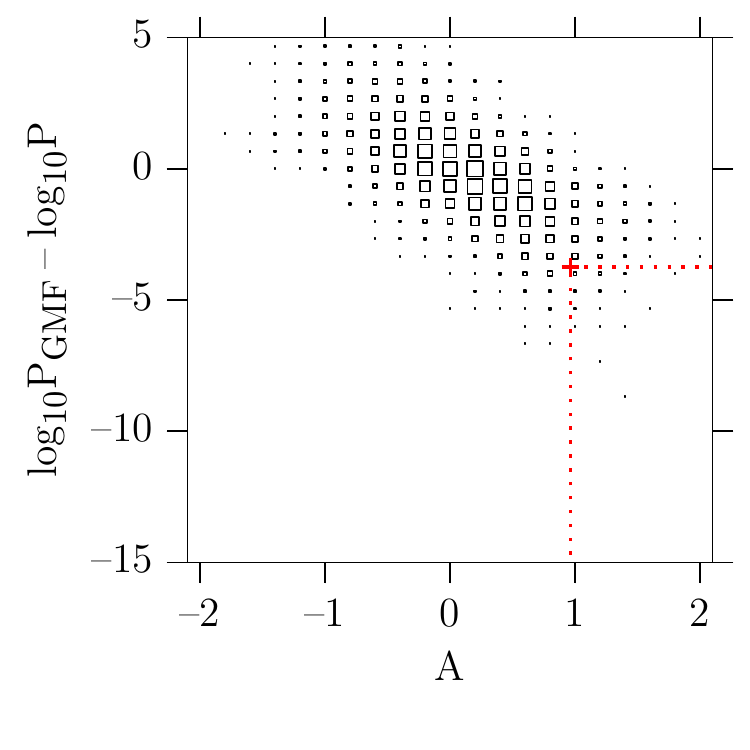}
}
\Block{d)}{
    %\vspace{-5ex}
    \centering
    \includegraphics[width=0.80\textwidth]{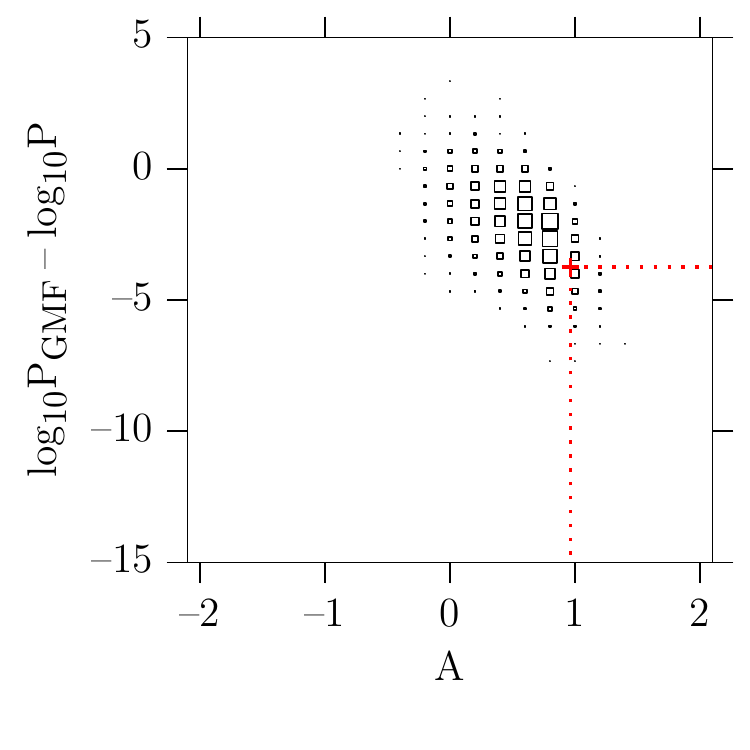}
}
    \\[5ex]
\caption{Improvement in the clustering strength versus the angular asymmetry using different analysis scenarios.
The red cross denotes the data analysis using in a,c,d) the regular field, and in b) a reverse field.
The box symbols denote the following variations:
a,b) random field directions and isotropic cosmic rays,
c) isotropic source directions,
d) source directions varied by $15$ deg.
Refer to text for details.
}
\label{fig:tests}
\end{figure}

In the previous section we performed the analysis on the simulated data set 
using the correct magnetic field and the correct set of AGN sources.
With $7\%$ signal protons in the data and $93\%$ background contributions, the two  
observables indicated overall improvements in terms of the angular
asymmetry $A=0.96$, and the clustering strength 
$\log_{10}(P_{\rm GMF})-\log_{10}(P)=-3.8$.
The results are visualized in Fig.\ref{fig:tests}a,c,d by the red cross.
\par
Below we investigate whether the same improvements can be obtained by chance.
In a first test we apply typical values for the expected deflections in the magnetic field; however,
we assume neither that the cosmic rays and AGNs are truly correlated,
nor that the above pattern of the magnetic field is correct.
\par
We use a simulation of $10,000$ event samples with the AGN directions,
isotropic cosmic ray arrival directions, and random patterns for the galactic 
magnetic field corrections.
The geometrical acceptance of the Pierre Auger Observatory is included as presented in \cite{Sommers2000}.
The distribution of the change in the clustering strength versus the angular asymmetry 
obtained from these simulations is shown in Fig.~\ref{fig:tests}a by the box symbols, 
and appears to be centered at zero with a slight anti-correlation in the two observables.
In $0.05\%$ of the events we find equal values or improvements in the two observables 
with respect to the values obtained in the data analysis.
\par
In a further test we use lenses produced for correcting deflections of antiprotons in the 
galactic magnetic field.
These lenses reverse the galactic magnetic field for protons.
Applying the reversed field in our data analysis instead of the correct field orientation, 
the angular asymmetry and the change in the clustering strength both disappear
(red cross in Fig.~\ref{fig:tests}b; the box symbols refer again to a simulation with random field 
directions, isotropic cosmic rays, and nominal AGN directions).
Thus, the above-mentioned second assumption on the validity of the field parameterizations can
be evaluated in comparison with analyses using random directions and field reversal.
\par
We also investigated the above-mentioned first assumption on angular correlations between 
AGNs and cosmic rays.
For this test we took the nominal field parameterization and the nominal cosmic ray arrival 
directions, but $22$ chance directions instead of the AGNs (box symbols in Fig.~\ref{fig:tests}c,
compared to the cosmic ray data analysis denoted by the red cross).
Also here the chance distribution is centered at zero, and is disjunct from the values 
obtained with the data.
A similar distribution is obtained when using isotropic cosmic rays, while keeping
the nominal field parameterization (JF12) and the nominal AGN directions (not shown here).
Thus, finding improvements at the level of our data analysis with arbitrary directions is unlikely,
implying that angular correlations of cosmic rays and sources can be evaluated.
\par
We also studied the influence of a potentially limited knowledge about the source directions.
In this test we first performed the analysis using the correct magnetic field and the 
nominal cosmic ray data set (red cross in Fig.~\ref{fig:tests}d).
We then varied the AGN positions within an angular uncertainty of $15$ deg.
Such variations could appear, e.g., when studying direct correlations of cosmic rays and cosmic neutrinos.
The box symbols show the results of the $10,000$ variations.
Even if the source directions are not perfectly known, an improvement in the two observables
owing to deflections in the galactic magnetic field can be observed.
\par
In applications of our method to measured cosmic rays data we therefore expect striking effects 
if the galactic magnetic field parameterization and the directions of the sources are correct.

\section{Conclusions}

In this contribution we presented a method for evaluating deflections of cosmic rays 
in the galactic magnetic field. 
By making use of a field parameterization derived from Faraday rotation and synchrotron 
emission measurements we calculate the expected arrival directions of protons and use
this for further analysis.
\par
To explore this method we introduced two observables designed for investigating cosmic
ray deflections, namely an angular asymmetry and a measure of clustering strength.
Applying the analysis to a simulated astrophysical scenario we demonstrated 
that the magnetic field corrections improve directional relations of cosmic rays and 
their sources if the correct field and source directions are used in the analysis.
This even works if the source directions are known with limited precision.
If incorrect assumptions on the field or an arbitrary set of sources are used, 
the improvements in the observables remain marginal.
In view of the galactic magnetic field parameterizations based on measurements
we thus expect that cosmic ray deflections in the galactic field can be verified experimentally.
\par
In general, and independently of the AGN application, our method of magnetic field corrections enables 
studies of directional correlations between cosmic rays and messenger particles suitable to 
indicate directions of cosmic ray sources.

\section*{Acknowledgments}

We are grateful for financial support to the Ministerium f\"ur Innovation, Wissenschaft und Forschung des 
Landes Nordrhein-Westfalen and to the Bundesministerium f\"ur Bildung und Forschung (BMBF).

\newpage

\end{document}